\documentclass[a4paper,11pt]{article}
\usepackage{pos}
\usepackage{wrapfig}
\usepackage{setspace}
\usepackage{enumitem}
\usepackage{multicol}
\newcommand{\Xmax}{X_\mathrm{max}}
\newcommand{\fpp}{f^\mathrm{pp}_{19}}
\newcommand{\deltaXmax}{\delta X_\mathrm{max}}

\title{Measurement of the Inelastic Proton-Proton Cross-Section at $\sqrt{s} \geq 40$ TeV Using the Hybrid Data of the Pierre Auger Observatory}
\ShortTitle{Measurement of the Proton-Proton Cross-Section}

\author*[a]{Olena Tkachenko}
\onbehalf{for the Pierre Auger Collaboration$^b$}

\affiliation[a]{Institute of Physics of the Czech Academy of Sciences, Na Slovance 1999/2, 182 00 Prague, Czech Republic}

\affiliation[b]{Observatorio Pierre Auger, Av.\ San Mart{\'\i}n Norte 304, 5613 Malarg\"ue, Argentina\\
Full author list: \normalfont{\url{https://www.auger.org/archive/authors_icrc_2025.html}}}

\emailAdd{spokespersons@auger.org}

\abstract{Measuring proton-proton interaction cross-sections at center-of-mass energies above 40 TeV remains a significant challenge in particle physics. The Pierre Auger Observatory provides a unique opportunity to study the interactions at the highest energies through the distribution of the depth of maximum shower development ($\Xmax$) observed by its Fluorescence Detector. In previous studies, the determination of the interaction cross-section at ultrahigh energies has relied on the assumption that the tail of the $\Xmax$ distribution is proton-dominated, which restricts the analysis to a limited energy range below the ankle and introduces related systematic uncertainties. In this contribution, we adopt a novel method for the simultaneous estimation of the proton-proton interaction cross-section and the primary cosmic-ray mass composition using data from the Pierre Auger Observatory, avoiding assumptions about one quantity to infer the other and thus improving the accuracy and robustness of our analysis. In addition, a systematic shift in the \(X_{\text{max}}\) scale is fitted to account for both experimental uncertainties and theoretical constraints on the modeling of particle interactions. The obtained results are consistent with previous analyses and provide additional constraints on hadronic interaction models. The measured proton-proton inelastic cross-section at ultra-high energies agrees well with extrapolations of accelerator data. The inferred cosmic-ray composition and the $\Xmax$-scale shift are also compatible with previous estimates.}

\ConferenceLogo{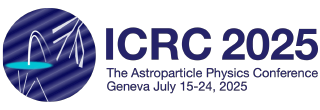}

\FullConference{39th International Cosmic Ray Conference (ICRC2025)\\
 15–24 July 2025\\
Geneva, Switzerland\\}

\begin{document}
\maketitle
\section{Introduction}

A key observable in the studies of ultrahigh-energy cosmic rays (UHECRs) is the atmospheric depth at which the air shower reaches its maximum development, denoted $\Xmax$. This quantity is highly sensitive to both the mass composition of the primary cosmic rays and the hadronic interaction properties of the shower development. Specifically, $\Xmax$ scales with the logarithm of the mass of the primary particle, providing information on its nuclear mass while also reflecting the depth of the first atmospheric interaction, which depends on the interaction cross-section of the incoming particle with air nuclei. However, at ultrahigh energies, neither the mass composition nor the hadronic cross-sections are directly or precisely measured, and estimating one often requires assumptions about the other. This mutual dependence is further complicated by the fact that the $\Xmax$ scale itself is not well constrained by current hadronic interaction models~\cite{PhysRevD.109.102001}, introducing additional uncertainty into the interpretation of both observables.

Addressing these interconnected uncertainties is critical. Studies of the mass composition help constrain the possible astrophysical sources of UHECRs, while probing hadronic interactions at these extreme energies offers insight into soft QCD processes that are well beyond the reach of human-made accelerators. Traditionally, measurements of the proton-air or proton-proton cross-sections, which is central to constraining hadronic models, have relied on the assumption that the tail of the $\Xmax$ distribution is dominated by protons, with a given helium contribution treated as a systematic uncertainty (typically limited to $\sim$25\%)~\cite{auger-2012, ulrich-2015}. Simultaneously, predictions of the mass composition rely on the same interaction models, making accurate cross-section estimates essential.

In this contribution, we refine the measurement of the proton-proton interaction cross-section through a simultaneous fit of the cross-section, mass composition, and $\Xmax$ scale using the data collected by the Fluorescence Detector (FD) of the Pierre Auger Observatory between December~1,~2004, and December 31, 2021~\cite{fitoussi-2024}. With this improved approach, we address the drawbacks of separate analyses and achieve a consistent estimation of the mass composition, the proton-proton interaction cross-section, and the $X_{\text{max}}$ scale while reducing reliance on the assumptions used in standard analyses.

\section{Method}

We determine the proton-proton interaction cross-section and UHECR mass composition through a simultaneous fit to the full \(X_{\text{max}}\) distribution~\cite{icrc2021-xsec, uhecr2024-xsec}, using predictions from air shower simulations with modified hadronic interaction properties. 
This approach builds on the standard mass composition fit~\cite{auger-composition2014}, using the same data selection procedure, with updates to both the selection and the reconstruction of the longitudinal air shower profile detailed in~\cite{fitoussi-2024, bellido-2023}. A further extension to the analysis is made by incorporating model predictions with modified interaction cross-sections. Additionally, it allows the \(X_{\text{max}}\) scale to vary freely, accounting for associated experimental and theoretical uncertainties. The proton-proton cross-section is rescaled by introducing an energy-dependent factor $f_\mathrm{lg E_1}$, following the approach used in the original cross-section analysis by the Pierre Auger Collaboration~\cite{auger-2012, ulrich-2011}. The modifications are implemented within the \textsc{Sibyll}~2.3\textnormal{d} model~\cite{riehn-2020} using the \textsc{CONEX} air shower simulation code~\cite{bergmann-2007}. Due to the \(E/A\) scaling characteristic of the superposition model, the effective onset of cross-section modifications is shifted to higher energies for nuclei with larger mass numbers. Consistently with the original analysis, the reference energy \(E_1\) is fixed at \(10^{19}\,\text{eV}\), which defines the energy at which the rescaling factor reaches the constant parameter value \(\fpp\). The threshold energy, above which the modifications in the cross-section are implemented, is set at the LHC center-of-mass energy. The nucleus-air cross-sections, calculated within the modified hadronic interaction model, are obtained via Glauber theory from the modified proton-proton cross-sections~\cite{glauber}.

First, $\Xmax$ distribution templates, i.e., precomputed simulated distributions used for comparison with data, are generated for discrete values of the rescaling factor $\fpp$, ranging from 0.2 to 2.0 in steps of 0.1. For each value of \(\fpp\), an \(\Xmax\) template that includes the detector acceptance and resolution effects is produced. Independently, a shift in \(\Xmax\), denoted \(\deltaXmax\), is applied to the data. This results in a two-dimensional scan over \((\fpp, \deltaXmax)\), yielding a best-fit mass composition for each parameter pair. The quality of the fit is assessed using the deviance, defined as the logarithm of the Poisson likelihood, which is approximately \(\chi^2\)-distributed. We identify the values of \(\fpp\) and \(\deltaXmax\) that minimize the total \(\chi^2\) across all energy bins, assuming both parameters are energy-independent. This analysis assumes a mass-independent \(X_{\text{max}}\)-scale shift. A fit including a \(\ln A\)-dependent term showed no significant mass dependence, with the coefficient consistent with zero.

\begin{figure}
\centering
\includegraphics[width=0.41\textwidth]{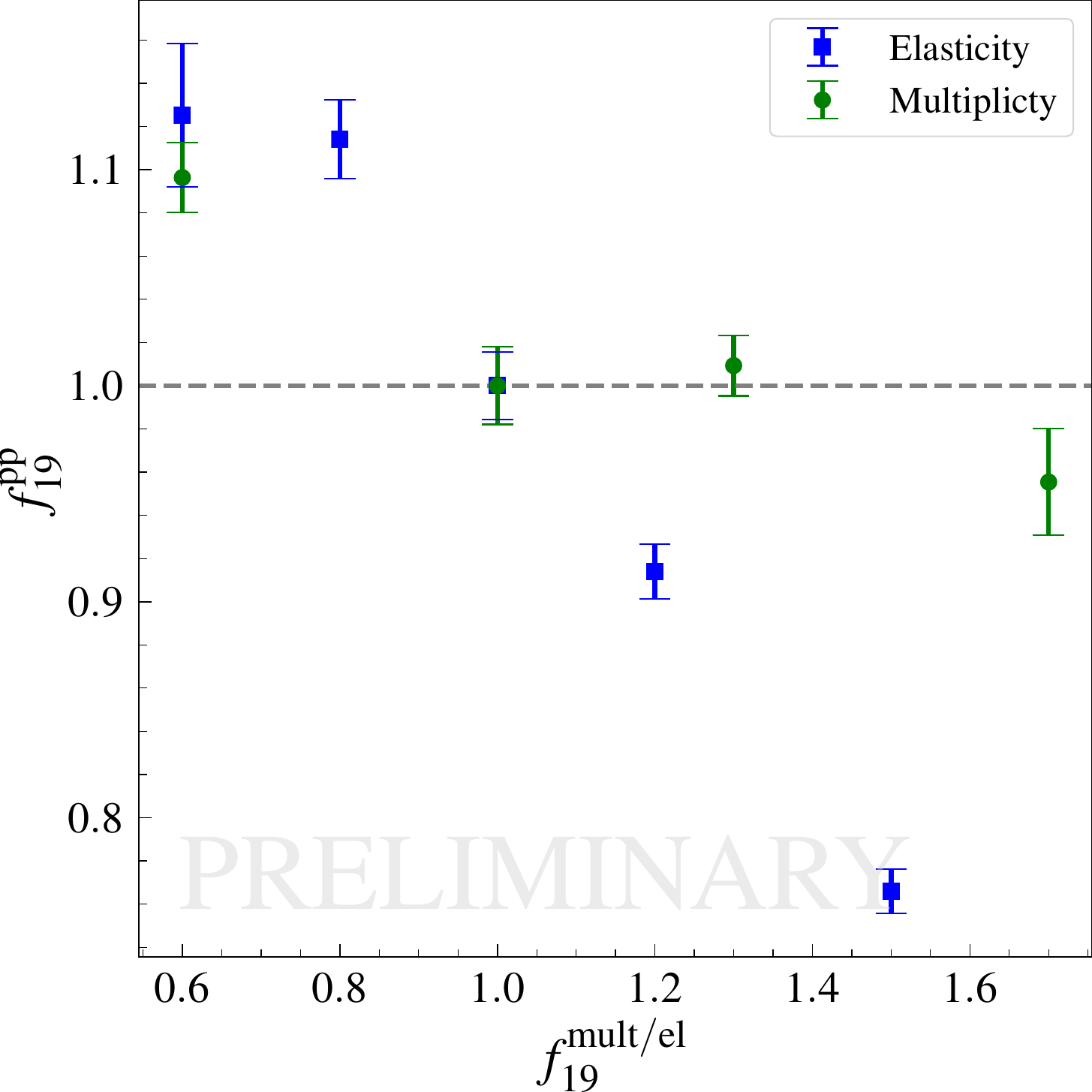}
\hspace{1.2cm}
\includegraphics[width=0.41\textwidth]{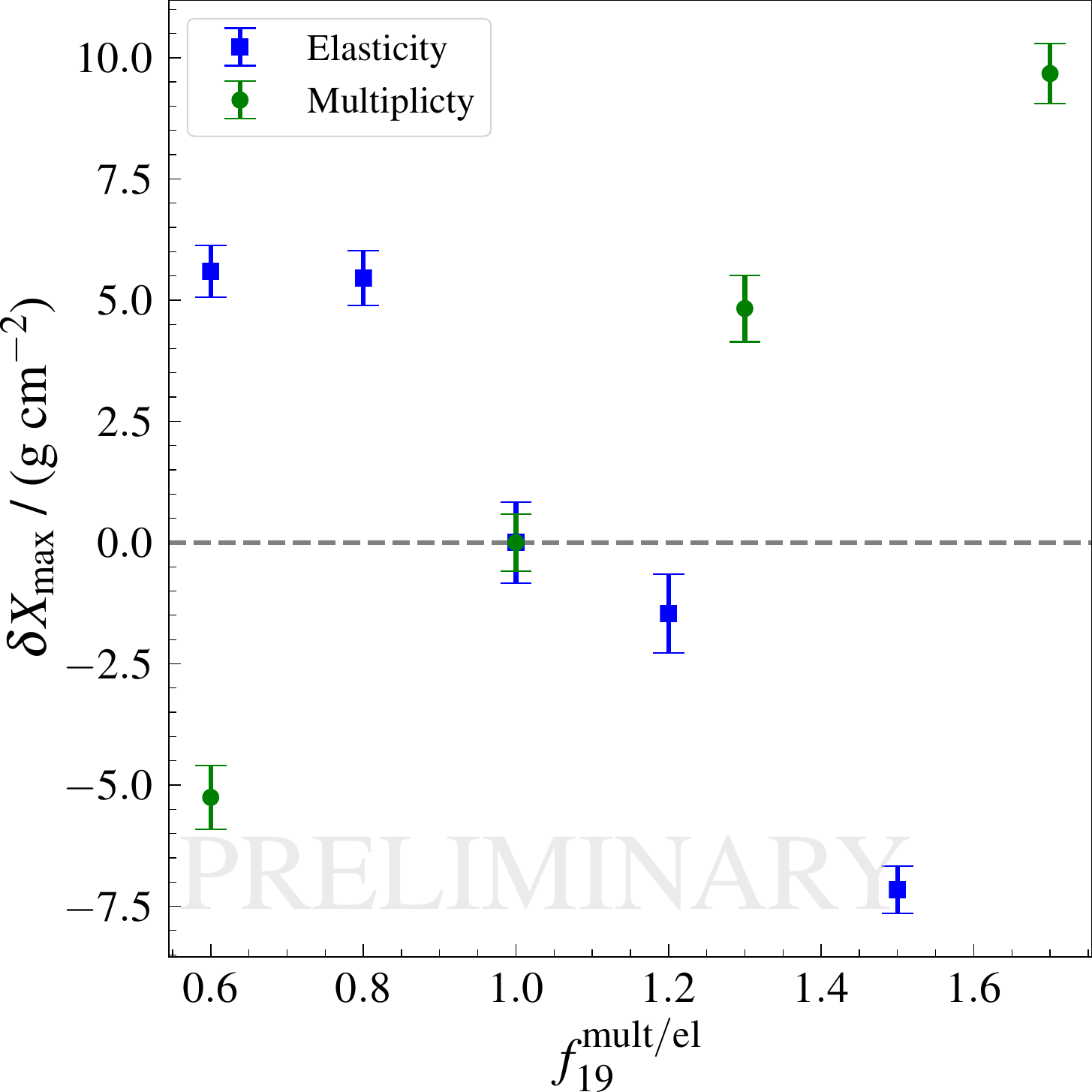}
\caption{Biases in the fitted cross-section rescaling factor \(\fpp\) (left) and shift in the \(\Xmax\) scale \(\deltaXmax\) (right) resulting from deviations in elasticity and multiplicity from the values assumed in the model prediction.}
\label{varied-elmul-f19}
\end{figure}

\section{Uncertainties and Biases}
Beyond statistical uncertainties, we account for relevant systematic effects, including those related to the energy scale, detector response, and hadronic interaction properties, specifically multiplicity and elasticity, as well as energy-dependent \(X_{\text{max}}\) systematics. Detector and energy scale systematics are evaluated via Monte Carlo simulations incorporating the detector response of the Pierre Auger Observatory~\cite{auger-2021-realMC}, following the standard \(X_{\text{max}}\) mass composition analysis~\cite{auger-composition2014}. 
Uncertainties from extrapolating accelerator measurements of elasticity and multiplicity are assessed using \textsc{CONEX} simulations, with these parameters being modified by following the same approach as for the cross-section variations~\cite{ulrich-2011, Ebr-MOCHI}. The modified simulations are then refitted. In Figure~\ref{varied-elmul-f19}, the resulting biases in the estimated cross-section and \(X_{\text{max}}\) scale are shown. The largest deviations observed in \(\fpp\), \(\delta X_{\text{max}}\), and the associated fractions are reported as systematic uncertainties.

The \(X_{\text{max}}\) scale is affected by several systematic sources, primarily reconstruction effects at low energies and atmospheric variations at high energies~\cite{PRD14-I}. Although these contributions combine to produce a nearly constant total uncertainty, their energy dependence can introduce both a global offset and an energy-dependent bias. To incorporate these dependencies, we generate multiple realizations of correlated \(X_{\text{max}}\) shifts across energy bins by random sampling within the uncertainty ranges of each source, preserving their energy correlations.

A potentially overlooked factor in \(X_{\text{max}}\)-based analyses is a reconstruction bias that depends on the \(X_{\text{max}}\) value itself. Monte Carlo simulations show that both the mean and width of the difference between generated and reconstructed \(X_{\text{max}}\) increase with larger \(X_{\text{max}}\), biasing the distribution tail by up to \(15\, \mathrm{g\,cm}^{-2}\). Since the tail is sensitive to interaction properties, this bias can significantly affect the fitted cross-section, while the central part remains stable with minimal bias. Because true \(X_{\text{max}}\) values are unknown, the correction is applied to simulation templates by adjusting the double-Gaussian resolution parameters. The bias is modeled as a cubic function of \(X_{\text{max}}\) and energy. It mainly affects the tail, with negligible impact on the mass composition, but it can bias the cross-section estimate by up to 10\%, thus, correcting for it is essential for accurate cross-section determination. It is important to note that this bias, if present, also impacts the previously adopted method of measuring cross-sections from the tail of the \(X_{\text{max}}\) distribution, with its extent depending on the profile reconstruction method. Neglecting this bias leads to an underestimation of \(\Lambda_\eta\), which consequently causes an overestimation of the cross-section. Therefore, applying the \(X_{\text{max}}\)-dependent correction is essential for both approaches.

\begin{figure}
\includegraphics[width=0.45\textwidth]{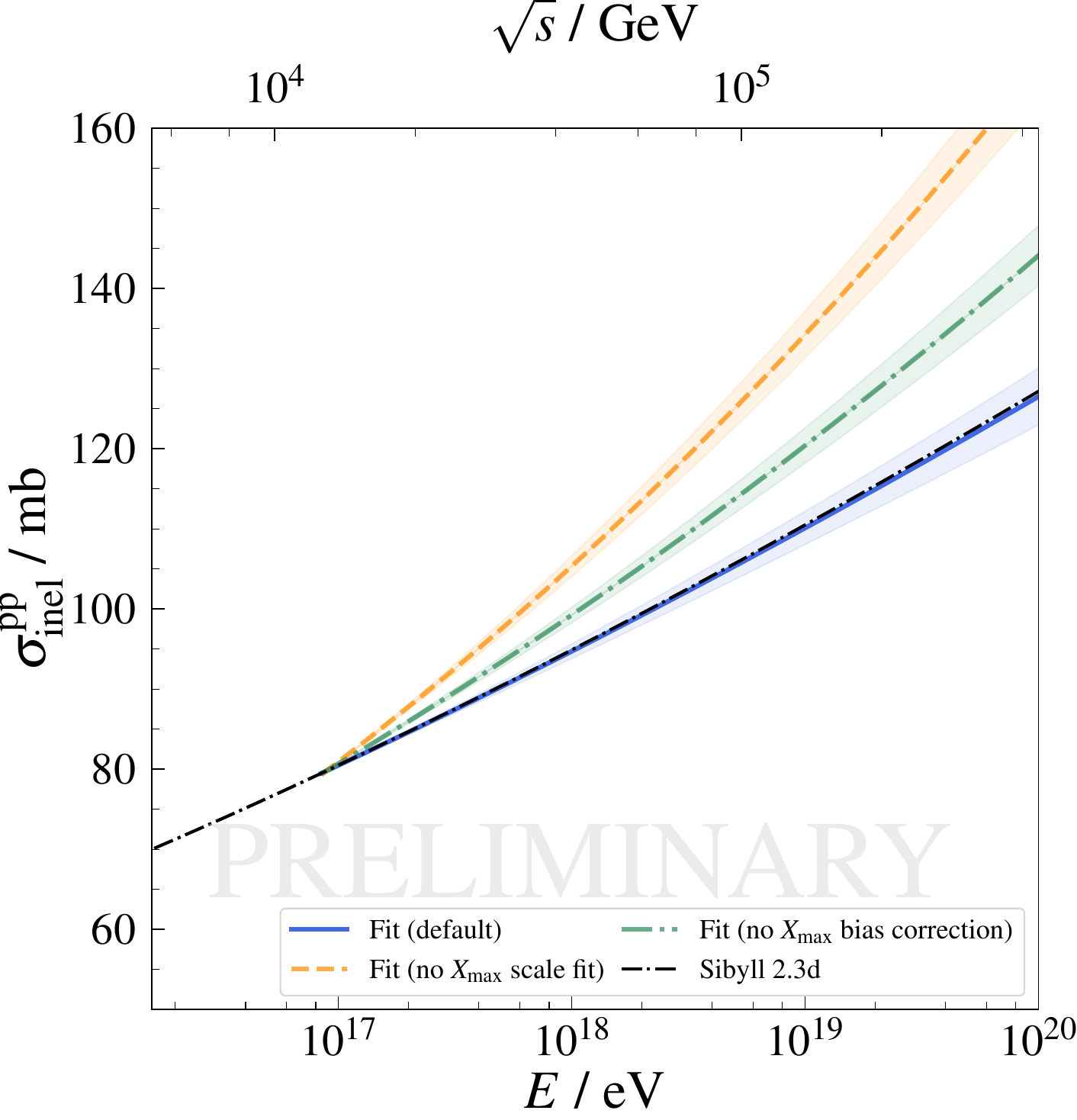}
\hspace{0.5cm}
\includegraphics[width=0.54\textwidth]{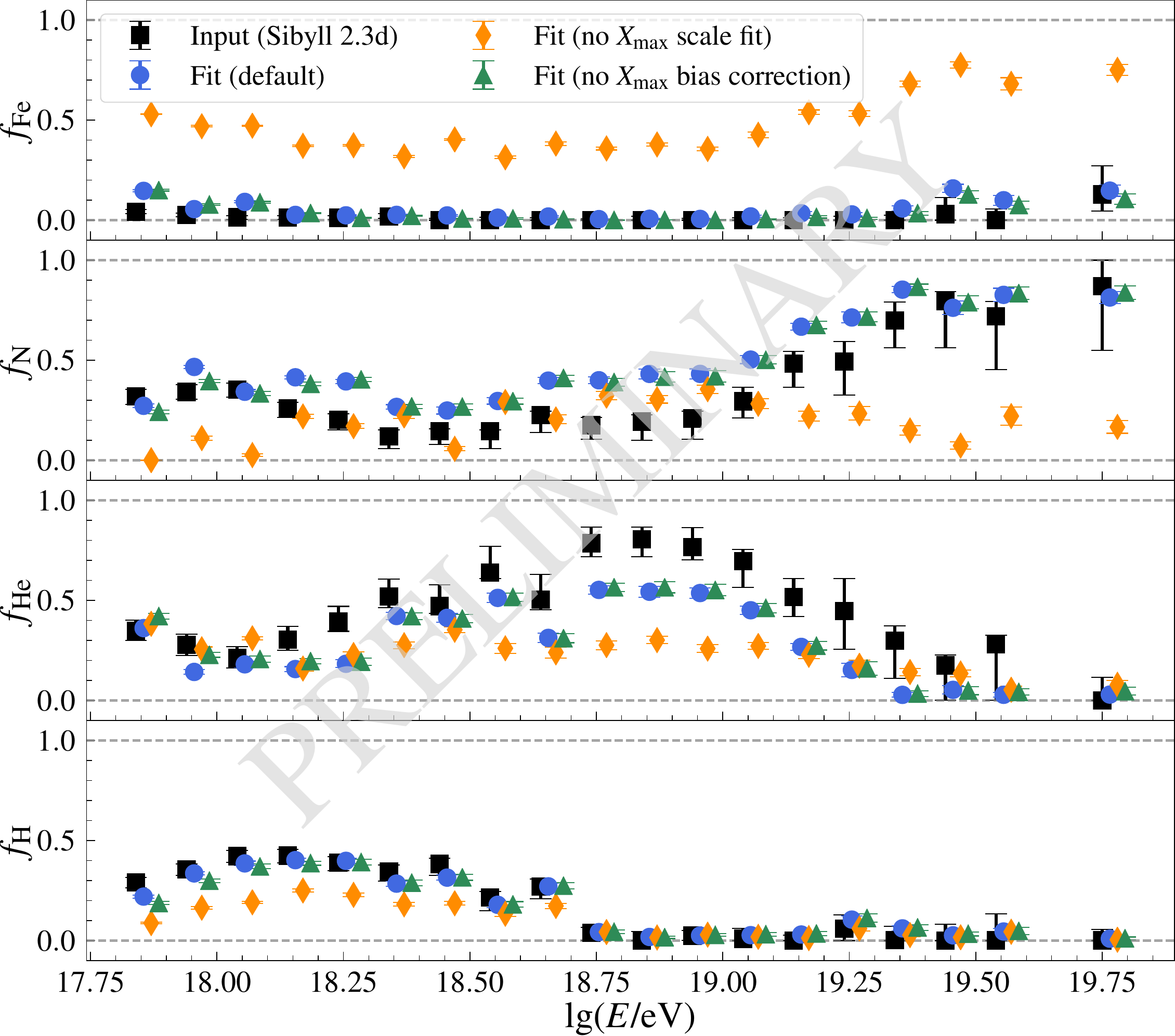}
\caption{Validation of the fitting procedure on the simulations. Left: Proton-proton cross-section. Right: Fitted composition fractions. To illustrate the impact of the discussed biases, results are additionally presented for fits performed with a fixed \(X_{\text{max}}\) scale and without applying the correction for the \(X_{\text{max}}\) bias.}
\label{Fig:Xmax-biases-tests}
\end{figure}

Unlike the standard method, which infers the interaction cross-section from the tail of the \(X_{\text{max}}\) distribution, the simultaneous fit of cross-section and mass composition yields an almost unbiased estimate that is largely independent of the helium fraction or overall mass composition, as demonstrated previously~\cite{uhecr2024-xsec}. In Figure~\ref{Fig:Xmax-biases-tests}, we show the fit results for simulations generated using the composition observed in the Pierre Auger Observatory data and incorporating the detector response. A negative \(30\,\mathrm{g\,cm}^{-2}\) shift in the \(X_{\text{max}}\) scale was included to test the impact of an unaccounted deviation from the \textsc{Sibyll}~2.3\textnormal{d} model. For comparison, the fit without correcting the \(X_{\text{max}}\)-dependent bias is also shown. When both the scale shift and bias correction are applied, the reconstructed values agree well with the simulated inputs, including composition, \(X_{\text{max}}\) shift, and cross-section. Small residual bias of approximately \(6\,\mathrm{g\,cm}^{-2}\) in \(X_{\text{max}}\) scale and an offset in composition are corrected in the fitted fractions and accounted for as systematic uncertainties in \(\fpp\) and \(\delta X_{\text{max}}\). If the \(X_{\text{max}}\) shift is omitted from the fit, the cross-section is overestimated, and the composition is biased toward lighter elements, as the fit compensates for the differing \(X_{\text{max}}\) scale by adjusting the shape of the distribution.

\section{Results}
We applied the fit procedure described above to the most recent data of the Pierre AugerObservatory, obtaining the cross-section, mass composition, and shift in the \(X_{\text{max}}\) scale. These results are then compared to those from the standard mass composition fit, the cross-section derived from the tail of the \(X_{\text{max}}\) distribution, and the latest Auger study evaluating hadronic model predictions for the \(X_{\text{max}}\) scale.

\begin{wrapfigure}{r}{0.55\textwidth}
  \vspace{-0.7cm}
  \begin{center}
   \includegraphics[width=0.5\textwidth]{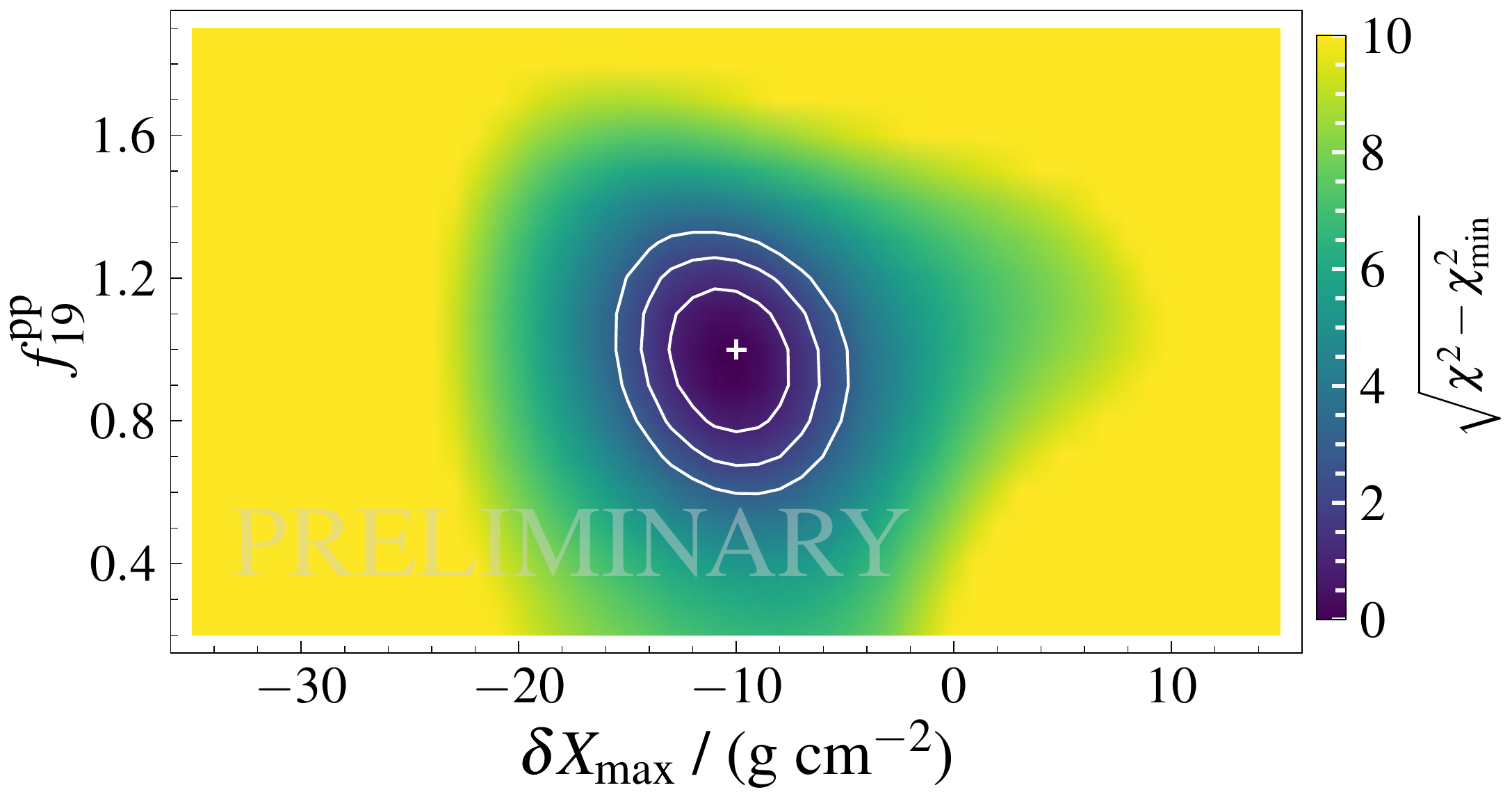}
   \caption{\(\chi^2\) contour of the fit, with the \(x\)-axis showing the \(X_{\text{max}}\)-scale shift and the \(y\)-axis the cross-section rescaling factor. The color scale represents the difference between the \(\chi^2\) value at each parameter point and the minimum \(\chi^2\) value across the full parameter space, with contours indicating the confidence levels.}
   \label{2d-scan}
  \end{center}
  \vspace{-0.65cm}
\end{wrapfigure}
\begin{figure}[t]
\centering
\includegraphics[width=0.73\textwidth]{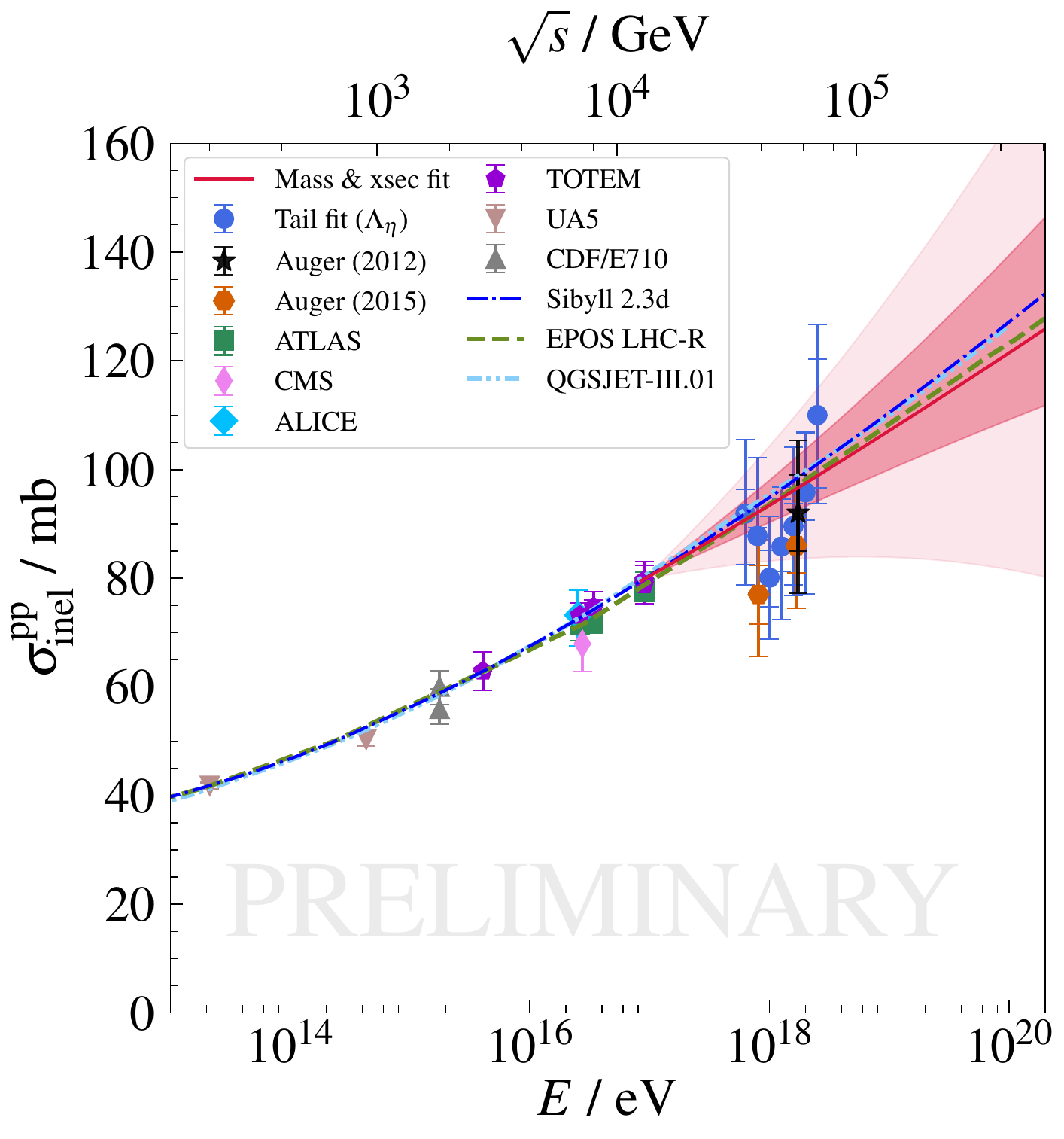}
\caption{The inelastic proton-proton cross-section estimated from Pierre Auger Observatory data with a simultaneous fit of mass composition, cross-section and a shift in \( X_{\text{max}} \) scale. The results are compared with predictions from contemporary hadronic interaction models, accelerator data, and measurements derived from the tail of the \( X_\mathrm{max} \) distribution. The darker shaded band indicates the statistical uncertainty, while the lighter band represents the total uncertainty.}
\label{Fig:cross-sec}
\end{figure}
\begin{figure}
\centering
\includegraphics[width=0.92\textwidth]{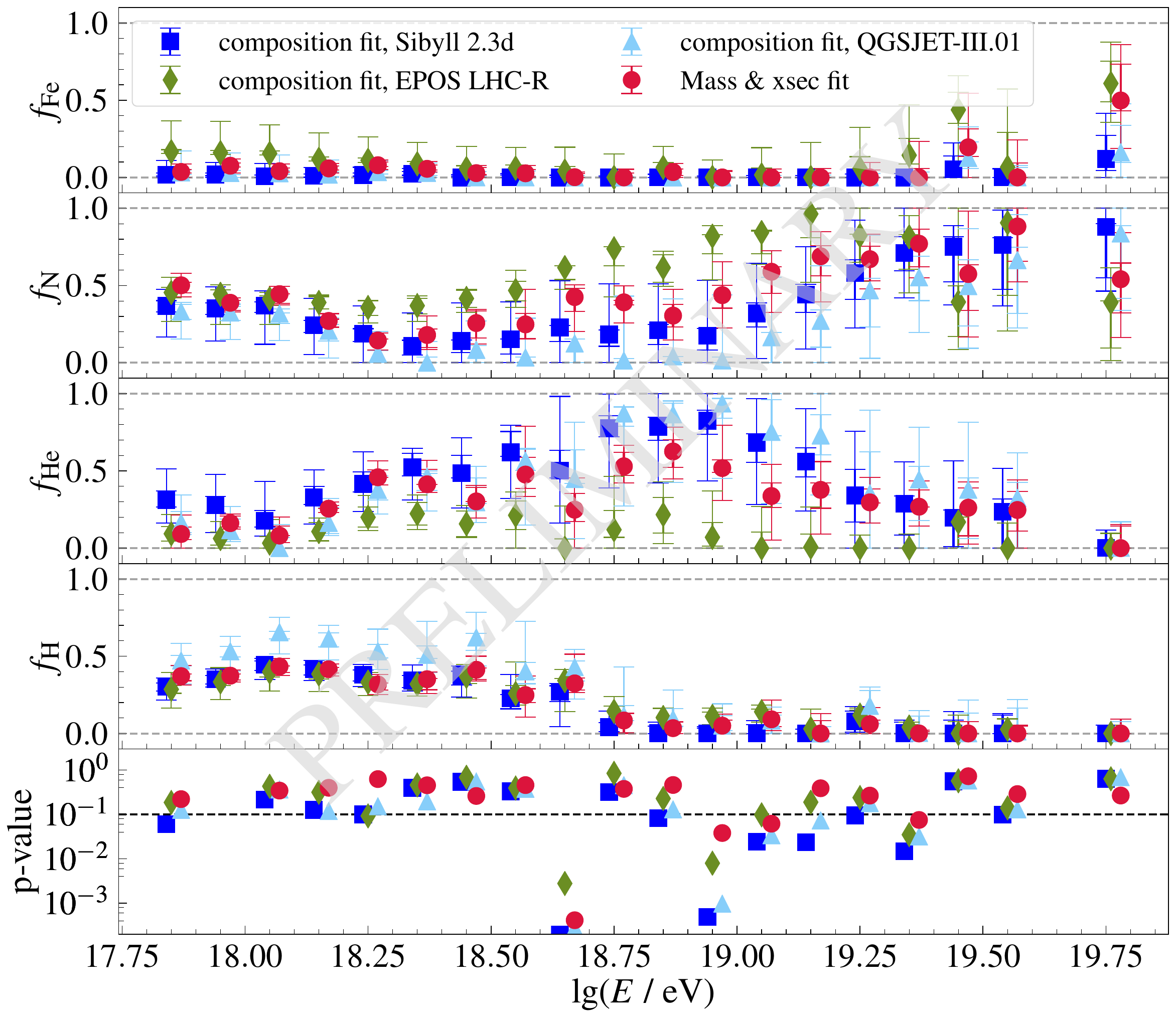}
\caption{The mass composition estimated from Pierre Auger Observatory data with a simultaneous fit of mass composition, cross-section and a shift in \( X_{\text{max}} \) scale. The results are compared with predictions from the composition-only fit using the predictions of the unmodified \textsc{Sibyll}~2.3\textnormal{d}, \textsc{EPOS~LHC‑R}, and \textsc{QGSJET-III.01} interaction models. The inner caps represent statistical uncertainties, while the outer caps indicate the total uncertainties, including both statistical and systematic contributions. The bottom panel displays the quality of the fit.}
\label{Fig:mass-composition}
\end{figure}

In Figure~\ref{2d-scan}, we show the two-dimensional scan over the cross-section rescaling factor and the shift in \(X_{\text{max}}\), from which the best-fit values of both parameters are determined. The displayed contours correspond to 68\% confidence intervals, obtained using a profile likelihood method based on the \(\chi^2 - \chi^2_\text{min} = 1\) criterion. In Figure~\ref{Fig:cross-sec}, the estimated cross-sections are presented, while in Figure~\ref{Fig:mass-composition}, we show the corresponding mass composition. Both observables are compared to the predictions from the original (unmodified) \textsc{Sibyll}~2.3\textnormal{d} model, as well as to those from the recently released \textsc{EPOS~LHC\mbox{-}R}~\cite{EPOSLHCR} and \textsc{QGSJET-III.01}~\cite{QGSJETIII} hadronic interaction models. The estimated rescaling factor is \( f^\mathrm{pp}_{19}~=~0.97^{+0.09}_{-0.07} \, \text{(stat.)}^{+0.24}_{-0.18} \, \text{(syst.)} \), and the corresponding shift in the \( X_{\text{max}} \) scale is \( \delta X_{\text{max}} = -10.3^{+1.7}_{-1.5} \, \text{(stat.)} ^{+13.6}_{-14.3} \, \text{(syst.)} \, \mathrm{g\,cm}^{-2} \). The quoted systematic uncertainties account for the components and biases discussed in the previous section. Although the measured cross-section and \( X_{\text{max}} \) values deviate from the predictions of the unmodified  \textsc{Sibyll}~2.3\textnormal{d} model, they remain consistent within the total uncertainty. Notably, the direction of the observed deviations aligns with the trend predicted by the \textsc{EPOS~LHC-R} model, which suggests a lower cross-section and a deeper \( X_{\text{max}} \) compared to  \textsc{Sibyll}~2.3\textnormal{d}. The comparison with the cross-sections derived from the tail of the \( X_{\text{max}} \) distribution shows good agreement in the energy range where the relative helium fraction remains sufficiently low for the tail method to provide reliable estimates.

The overall \( X_{\text{max}} \)-scale shift obtained here is about \(20\,\mathrm{g\,cm}^{-2}\) smaller than the one reported in the \((S(1000), X_{\text{max}})\) analysis~\cite{PhysRevD.109.102001}. However, it is worth noting that the results presented in these proceedings depend on the selected energy range. Restricting our fit to the same energy interval used in that study yields a shift of \( \delta X_{\text{max}} = -15.6^{+2.2}_{-3.2} \, \text{(stat.)}^{+13}_{-18} \, \text{(syst.)} \, \mathrm{g\,cm}^{-2} \), consistent within systematic uncertainties with the previously reported shift in the model predictions. The rescaling factor \( f^\mathrm{pp}_{19} = 0.74^{+0.14}_{-0.18} \, \text{(stat.)} \) suggests an additional negative shift and broader \( X_{\text{max}} \) distribution. Differences in energy range and analysis assumptions, however, complicate a direct comparison between the two analyses.

Relative to the composition‑only fit with \textsc{Sibyll}~2.3\textnormal{d}, the mass composition obtained from the simultaneous composition and cross‑section fit is heavier, yet still lighter than that predicted by \textsc{EPOS~LHC‑R}. In comparison with the unmodified \textsc{Sibyll}~2.3\textnormal{d} model, the proton and iron fractions remain nearly unchanged across most of the energy range, except at the highest energies where the iron fraction increases by up to 40\%. The nitrogen and helium fractions, on the other hand, show more noticeable deviations from the unmodified \textsc{Sibyll}~2.3\textnormal{d} model, with the nitrogen fraction increasing by up to 20\% in the intermediate energy range, accompanied by a corresponding decline in helium contamination. Overall, the new composition sits between the \textsc{Sibyll}~2.3\textnormal{d} and \textsc{EPOS~LHC‑R} predictions and remains compatible with both within the quoted total uncertainties. The improved agreement between the model and the data is reflected in the higher p-values, as shown in the bottom panel of Figure~\ref{Fig:mass-composition}.

\section{Conclusions}
In this work, we present a joint analysis of the cosmic-ray mass composition and the inelastic proton-proton cross-section using the full Phase~I dataset from the Pierre Auger Observatory. Both quantities are extracted via a simultaneous fit to the \(X_{\text{max}}\) distributions, with the \(X_{\text{max}}\) scale included as a free parameter to address uncertainties in detector response and hadronic interaction models. This integrated approach reduces dependence on fixed assumptions, such as a proton-dominated \(X_{\text{max}}\) tail or fixed model predictions, enhancing the internal consistency of the analysis. Compared to previous measurements, the statistical uncertainty on the proton-proton cross-section has decreased by about 20\%, thanks to the larger dataset and methodological improvements. Systematic uncertainties are also lowered by incorporating dominant error sources directly into the fit. The applied method additionally improves the reliability of mass fraction estimates by explicitly including the proton-proton cross-section, which significantly contributes to model-related uncertainties, within the simultaneous fit. These advances enable a more precise and robust characterization of mass composition and cross-section, allowing for the extension of cross-section measurements to higher energies with reduced bias from heavier primaries, such as helium. The results obtained in this work suggest a somewhat heavier mass composition, a slightly lower proton-proton cross-section, and a small negative shift in the data relative to model predictions. In summary, the estimated proton-proton interaction cross-section, as well as the mass composition and the \(X_{\text{max}}\) scale, remain consistent with previous measurements and theoretical model extrapolations within the quoted systematic uncertainties. This agreement supports the validity of the previous and current analyses while acknowledging the existing uncertainties and limitations inherent in the data and models.

\begin{multicols}{2}
\begin{spacing}{0.8}
\end{spacing}
\end{multicols}
\newpage
\begin{center}
\par\noindent
\textbf{The Pierre Auger Collaboration}
\end{center}
\begin{wrapfigure}[9]{l}{0.12\linewidth}
\vspace{-2.9ex}
\includegraphics[width=0.98\linewidth]{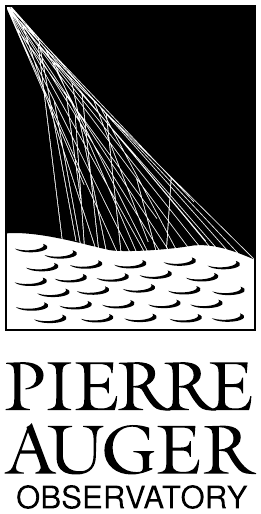}
\end{wrapfigure}
\begin{sloppypar}\noindent
% created on 2025-06-06
A.~Abdul Halim$^{13}$,
P.~Abreu$^{70}$,
M.~Aglietta$^{53,51}$,
I.~Allekotte$^{1}$,
K.~Almeida Cheminant$^{78,77}$,
A.~Almela$^{7,12}$,
R.~Aloisio$^{44,45}$,
J.~Alvarez-Mu\~niz$^{76}$,
A.~Ambrosone$^{44}$,
J.~Ammerman Yebra$^{76}$
% ...

\end{sloppypar}

\vspace{1ex}
\begin{center}
\rule{0.1\columnwidth}{0.5pt}
\raisebox{-0.4ex}{\scriptsize$\bullet$}
\rule{0.1\columnwidth}{0.5pt}
\end{center}

\vspace{1ex}
% created on 2025-06-06
% needs \usepackage{enumitem}
\begin{description}[labelsep=0.2em,align=right,labelwidth=0.7em,labelindent=0em,leftmargin=2em,noitemsep,before={\renewcommand\makelabel[1]{##1 }}]
\item[$^{1}$] Centro At\'omico Bariloche and Instituto Balseiro (CNEA-UNCuyo-CONICET), San Carlos de Bariloche, Argentina
\item[$^{2}$] Departamento de F\'\i{}sica and Departamento de Ciencias de la Atm\'osfera y los Oc\'eanos, FCEyN, Universidad de Buenos Aires and CONICET, Buenos Aires, Argentina
\item[$^{3}$] IFLP, Universidad Nacional de La Plata and CONICET, La Plata, Argentina
\item[$^{4}$] Instituto de Astronom\'\i{}a y F\'\i{}sica del Espacio (IAFE, CONICET-UBA), Buenos Aires, Argentina
\item[$^{5}$] Instituto de F\'\i{}sica de Rosario (IFIR) -- CONICET/U.N.R.\ and Facultad de Ciencias Bioqu\'\i{}micas y Farmac\'euticas U.N.R., Rosario, Argentina
\item[$^{6}$] Instituto de Tecnolog\'\i{}as en Detecci\'on y Astropart\'\i{}culas (CNEA, CONICET, UNSAM), and Universidad Tecnol\'ogica Nacional -- Facultad Regional Mendoza (CONICET/CNEA), Mendoza, Argentina
\item[$^{7}$] Instituto de Tecnolog\'\i{}as en Detecci\'on y Astropart\'\i{}culas (CNEA, CONICET, UNSAM), Buenos Aires, Argentina
\item[$^{8}$] International Center of Advanced Studies and Instituto de Ciencias F\'\i{}sicas, ECyT-UNSAM and CONICET, Campus Miguelete -- San Mart\'\i{}n, Buenos Aires, Argentina
\item[$^{9}$] Laboratorio Atm\'osfera -- Departamento de Investigaciones en L\'aseres y sus Aplicaciones -- UNIDEF (CITEDEF-CONICET), Argentina
\item[$^{10}$] Observatorio Pierre Auger, Malarg\"ue, Argentina
\item[$^{11}$] Observatorio Pierre Auger and Comisi\'on Nacional de Energ\'\i{}a At\'omica, Malarg\"ue, Argentina
\item[$^{12}$] Universidad Tecnol\'ogica Nacional -- Facultad Regional Buenos Aires, Buenos Aires, Argentina
\item[$^{13}$] University of Adelaide, Adelaide, S.A., Australia
\item[$^{14}$] Universit\'e Libre de Bruxelles (ULB), Brussels, Belgium
\item[$^{15}$] Vrije Universiteit Brussels, Brussels, Belgium
\item[$^{16}$] Centro Brasileiro de Pesquisas Fisicas, Rio de Janeiro, RJ, Brazil
\item[$^{17}$] Centro Federal de Educa\c{c}\~ao Tecnol\'ogica Celso Suckow da Fonseca, Petropolis, Brazil
\item[$^{18}$] Instituto Federal de Educa\c{c}\~ao, Ci\^encia e Tecnologia do Rio de Janeiro (IFRJ), Brazil
\item[$^{19}$] Universidade de S\~ao Paulo, Escola de Engenharia de Lorena, Lorena, SP, Brazil
\item[$^{20}$] Universidade de S\~ao Paulo, Instituto de F\'\i{}sica de S\~ao Carlos, S\~ao Carlos, SP, Brazil
\item[$^{21}$] Universidade de S\~ao Paulo, Instituto de F\'\i{}sica, S\~ao Paulo, SP, Brazil
\item[$^{22}$] Universidade Estadual de Campinas (UNICAMP), IFGW, Campinas, SP, Brazil
\item[$^{23}$] Universidade Estadual de Feira de Santana, Feira de Santana, Brazil
\item[$^{24}$] Universidade Federal de Campina Grande, Centro de Ciencias e Tecnologia, Campina Grande, Brazil
\item[$^{25}$] Universidade Federal do ABC, Santo Andr\'e, SP, Brazil
\item[$^{26}$] Universidade Federal do Paran\'a, Setor Palotina, Palotina, Brazil
\item[$^{27}$] Universidade Federal do Rio de Janeiro, Instituto de F\'\i{}sica, Rio de Janeiro, RJ, Brazil
\item[$^{28}$] Universidad de Medell\'\i{}n, Medell\'\i{}n, Colombia
\item[$^{29}$] Universidad Industrial de Santander, Bucaramanga, Colombia
\item[$^{30}$] Charles University, Faculty of Mathematics and Physics, Institute of Particle and Nuclear Physics, Prague, Czech Republic
\item[$^{31}$] Institute of Physics of the Czech Academy of Sciences, Prague, Czech Republic
\item[$^{32}$] Palacky University, Olomouc, Czech Republic
\item[$^{33}$] CNRS/IN2P3, IJCLab, Universit\'e Paris-Saclay, Orsay, France
\item[$^{34}$] Laboratoire de Physique Nucl\'eaire et de Hautes Energies (LPNHE), Sorbonne Universit\'e, Universit\'e de Paris, CNRS-IN2P3, Paris, France
\item[$^{35}$] Univ.\ Grenoble Alpes, CNRS, Grenoble Institute of Engineering Univ.\ Grenoble Alpes, LPSC-IN2P3, 38000 Grenoble, France
\item[$^{36}$] Universit\'e Paris-Saclay, CNRS/IN2P3, IJCLab, Orsay, France
\item[$^{37}$] Bergische Universit\"at Wuppertal, Department of Physics, Wuppertal, Germany
\item[$^{38}$] Karlsruhe Institute of Technology (KIT), Institute for Experimental Particle Physics, Karlsruhe, Germany
\item[$^{39}$] Karlsruhe Institute of Technology (KIT), Institut f\"ur Prozessdatenverarbeitung und Elektronik, Karlsruhe, Germany
\item[$^{40}$] Karlsruhe Institute of Technology (KIT), Institute for Astroparticle Physics, Karlsruhe, Germany
\item[$^{41}$] RWTH Aachen University, III.\ Physikalisches Institut A, Aachen, Germany
\item[$^{42}$] Universit\"at Hamburg, II.\ Institut f\"ur Theoretische Physik, Hamburg, Germany
\item[$^{43}$] Universit\"at Siegen, Department Physik -- Experimentelle Teilchenphysik, Siegen, Germany
\item[$^{44}$] Gran Sasso Science Institute, L'Aquila, Italy
\item[$^{45}$] INFN Laboratori Nazionali del Gran Sasso, Assergi (L'Aquila), Italy
\item[$^{46}$] INFN, Sezione di Catania, Catania, Italy
\item[$^{47}$] INFN, Sezione di Lecce, Lecce, Italy
\item[$^{48}$] INFN, Sezione di Milano, Milano, Italy
\item[$^{49}$] INFN, Sezione di Napoli, Napoli, Italy
\item[$^{50}$] INFN, Sezione di Roma ``Tor Vergata'', Roma, Italy
\item[$^{51}$] INFN, Sezione di Torino, Torino, Italy
\item[$^{52}$] Istituto di Astrofisica Spaziale e Fisica Cosmica di Palermo (INAF), Palermo, Italy
\item[$^{53}$] Osservatorio Astrofisico di Torino (INAF), Torino, Italy
\item[$^{54}$] Politecnico di Milano, Dipartimento di Scienze e Tecnologie Aerospaziali , Milano, Italy
\item[$^{55}$] Universit\`a del Salento, Dipartimento di Matematica e Fisica ``E.\ De Giorgi'', Lecce, Italy
\item[$^{56}$] Universit\`a dell'Aquila, Dipartimento di Scienze Fisiche e Chimiche, L'Aquila, Italy
\item[$^{57}$] Universit\`a di Catania, Dipartimento di Fisica e Astronomia ``Ettore Majorana``, Catania, Italy
\item[$^{58}$] Universit\`a di Milano, Dipartimento di Fisica, Milano, Italy
\item[$^{59}$] Universit\`a di Napoli ``Federico II'', Dipartimento di Fisica ``Ettore Pancini'', Napoli, Italy
\item[$^{60}$] Universit\`a di Palermo, Dipartimento di Fisica e Chimica ''E.\ Segr\`e'', Palermo, Italy
\item[$^{61}$] Universit\`a di Roma ``Tor Vergata'', Dipartimento di Fisica, Roma, Italy
\item[$^{62}$] Universit\`a Torino, Dipartimento di Fisica, Torino, Italy
\item[$^{63}$] Benem\'erita Universidad Aut\'onoma de Puebla, Puebla, M\'exico
\item[$^{64}$] Unidad Profesional Interdisciplinaria en Ingenier\'\i{}a y Tecnolog\'\i{}as Avanzadas del Instituto Polit\'ecnico Nacional (UPIITA-IPN), M\'exico, D.F., M\'exico
\item[$^{65}$] Universidad Aut\'onoma de Chiapas, Tuxtla Guti\'errez, Chiapas, M\'exico
\item[$^{66}$] Universidad Michoacana de San Nicol\'as de Hidalgo, Morelia, Michoac\'an, M\'exico
\item[$^{67}$] Universidad Nacional Aut\'onoma de M\'exico, M\'exico, D.F., M\'exico
\item[$^{68}$] Institute of Nuclear Physics PAN, Krakow, Poland
\item[$^{69}$] University of \L{}\'od\'z, Faculty of High-Energy Astrophysics,\L{}\'od\'z, Poland
\item[$^{70}$] Laborat\'orio de Instrumenta\c{c}\~ao e F\'\i{}sica Experimental de Part\'\i{}culas -- LIP and Instituto Superior T\'ecnico -- IST, Universidade de Lisboa -- UL, Lisboa, Portugal
\item[$^{71}$] ``Horia Hulubei'' National Institute for Physics and Nuclear Engineering, Bucharest-Magurele, Romania
\item[$^{72}$] Institute of Space Science, Bucharest-Magurele, Romania
\item[$^{73}$] Center for Astrophysics and Cosmology (CAC), University of Nova Gorica, Nova Gorica, Slovenia
\item[$^{74}$] Experimental Particle Physics Department, J.\ Stefan Institute, Ljubljana, Slovenia
\item[$^{75}$] Universidad de Granada and C.A.F.P.E., Granada, Spain
\item[$^{76}$] Instituto Galego de F\'\i{}sica de Altas Enerx\'\i{}as (IGFAE), Universidade de Santiago de Compostela, Santiago de Compostela, Spain
\item[$^{77}$] IMAPP, Radboud University Nijmegen, Nijmegen, The Netherlands
\item[$^{78}$] Nationaal Instituut voor Kernfysica en Hoge Energie Fysica (NIKHEF), Science Park, Amsterdam, The Netherlands
\item[$^{79}$] Stichting Astronomisch Onderzoek in Nederland (ASTRON), Dwingeloo, The Netherlands
\item[$^{80}$] Universiteit van Amsterdam, Faculty of Science, Amsterdam, The Netherlands
\item[$^{81}$] Case Western Reserve University, Cleveland, OH, USA
\item[$^{82}$] Colorado School of Mines, Golden, CO, USA
\item[$^{83}$] Department of Physics and Astronomy, Lehman College, City University of New York, Bronx, NY, USA
\item[$^{84}$] Michigan Technological University, Houghton, MI, USA
\item[$^{85}$] New York University, New York, NY, USA
\item[$^{86}$] University of Chicago, Enrico Fermi Institute, Chicago, IL, USA
\item[$^{87}$] University of Delaware, Department of Physics and Astronomy, Bartol Research Institute, Newark, DE, USA
\item[] -----
\item[$^{a}$] Max-Planck-Institut f\"ur Radioastronomie, Bonn, Germany
\item[$^{b}$] also at Kapteyn Institute, University of Groningen, Groningen, The Netherlands
\item[$^{c}$] School of Physics and Astronomy, University of Leeds, Leeds, United Kingdom
\item[$^{d}$] Fermi National Accelerator Laboratory, Fermilab, Batavia, IL, USA
\item[$^{e}$] Pennsylvania State University, University Park, PA, USA
\item[$^{f}$] Colorado State University, Fort Collins, CO, USA
\item[$^{g}$] Louisiana State University, Baton Rouge, LA, USA
\item[$^{h}$] now at Graduate School of Science, Osaka Metropolitan University, Osaka, Japan
\item[$^{i}$] Institut universitaire de France (IUF), France
\item[$^{j}$] now at Technische Universit\"at Dortmund and Ruhr-Universit\"at Bochum, Dortmund and Bochum, Germany
\end{description}

\vspace{1ex}
% created on 2025-06-06
\section*{Acknowledgments}

\begin{sloppypar}
The successful installation, commissioning, and operation of the Pierre
Auger Observatory would not have been possible without the strong
commitment and effort from the technical and administrative staff in
Malarg\"ue. We are very grateful to the following agencies and
organizations for financial support:
\end{sloppypar}

\begin{sloppypar}
Argentina -- Comisi\'on Nacional de Energ\'\i{}a At\'omica; Agencia Nacional de
Promoci\'on Cient\'\i{}fica y Tecnol\'ogica (ANPCyT); Consejo Nacional de
Investigaciones Cient\'\i{}ficas y T\'ecnicas (CONICET); Gobierno de la
Provincia de Mendoza; Municipalidad de Malarg\"ue; NDM Holdings and Valle
Las Le\~nas; in gratitude for their continuing cooperation over land
access; Australia -- the Australian Research Council; Belgium -- Fonds
de la Recherche Scientifique (FNRS); Research Foundation Flanders (FWO),
Marie Curie Action of the European Union Grant No.~101107047; Brazil --
Conselho Nacional de Desenvolvimento Cient\'\i{}fico e Tecnol\'ogico (CNPq);
Financiadora de Estudos e Projetos (FINEP); Funda\c{c}\~ao de Amparo \`a
Pesquisa do Estado de Rio de Janeiro (FAPERJ); S\~ao Paulo Research
Foundation (FAPESP) Grants No.~2019/10151-2, No.~2010/07359-6 and
No.~1999/05404-3; Minist\'erio da Ci\^encia, Tecnologia, Inova\c{c}\~oes e
Comunica\c{c}\~oes (MCTIC); Czech Republic -- GACR 24-13049S, CAS LQ100102401,
MEYS LM2023032, CZ.02.1.01/0.0/0.0/16{\textunderscore}013/0001402,
CZ.02.1.01/0.0/0.0/18{\textunderscore}046/0016010 and
CZ.02.1.01/0.0/0.0/17{\textunderscore}049/0008422 and CZ.02.01.01/00/22{\textunderscore}008/0004632;
France -- Centre de Calcul IN2P3/CNRS; Centre National de la Recherche
Scientifique (CNRS); Conseil R\'egional Ile-de-France; D\'epartement
Physique Nucl\'eaire et Corpusculaire (PNC-IN2P3/CNRS); D\'epartement
Sciences de l'Univers (SDU-INSU/CNRS); Institut Lagrange de Paris (ILP)
Grant No.~LABEX ANR-10-LABX-63 within the Investissements d'Avenir
Programme Grant No.~ANR-11-IDEX-0004-02; Germany -- Bundesministerium
f\"ur Bildung und Forschung (BMBF); Deutsche Forschungsgemeinschaft (DFG);
Finanzministerium Baden-W\"urttemberg; Helmholtz Alliance for
Astroparticle Physics (HAP); Helmholtz-Gemeinschaft Deutscher
Forschungszentren (HGF); Ministerium f\"ur Kultur und Wissenschaft des
Landes Nordrhein-Westfalen; Ministerium f\"ur Wissenschaft, Forschung und
Kunst des Landes Baden-W\"urttemberg; Italy -- Istituto Nazionale di
Fisica Nucleare (INFN); Istituto Nazionale di Astrofisica (INAF);
Ministero dell'Universit\`a e della Ricerca (MUR); CETEMPS Center of
Excellence; Ministero degli Affari Esteri (MAE), ICSC Centro Nazionale
di Ricerca in High Performance Computing, Big Data and Quantum
Computing, funded by European Union NextGenerationEU, reference code
CN{\textunderscore}00000013; M\'exico -- Consejo Nacional de Ciencia y Tecnolog\'\i{}a
(CONACYT) No.~167733; Universidad Nacional Aut\'onoma de M\'exico (UNAM);
PAPIIT DGAPA-UNAM; The Netherlands -- Ministry of Education, Culture and
Science; Netherlands Organisation for Scientific Research (NWO); Dutch
national e-infrastructure with the support of SURF Cooperative; Poland
-- Ministry of Education and Science, grants No.~DIR/WK/2018/11 and
2022/WK/12; National Science Centre, grants No.~2016/22/M/ST9/00198,
2016/23/B/ST9/01635, 2020/39/B/ST9/01398, and 2022/45/B/ST9/02163;
Portugal -- Portuguese national funds and FEDER funds within Programa
Operacional Factores de Competitividade through Funda\c{c}\~ao para a Ci\^encia
e a Tecnologia (COMPETE); Romania -- Ministry of Research, Innovation
and Digitization, CNCS-UEFISCDI, contract no.~30N/2023 under Romanian
National Core Program LAPLAS VII, grant no.~PN 23 21 01 02 and project
number PN-III-P1-1.1-TE-2021-0924/TE57/2022, within PNCDI III; Slovenia
-- Slovenian Research Agency, grants P1-0031, P1-0385, I0-0033, N1-0111;
Spain -- Ministerio de Ciencia e Innovaci\'on/Agencia Estatal de
Investigaci\'on (PID2019-105544GB-I00, PID2022-140510NB-I00 and
RYC2019-027017-I), Xunta de Galicia (CIGUS Network of Research Centers,
Consolidaci\'on 2021 GRC GI-2033, ED431C-2021/22 and ED431F-2022/15),
Junta de Andaluc\'\i{}a (SOMM17/6104/UGR and P18-FR-4314), and the European
Union (Marie Sklodowska-Curie 101065027 and ERDF); USA -- Department of
Energy, Contracts No.~DE-AC02-07CH11359, No.~DE-FR02-04ER41300,
No.~DE-FG02-99ER41107 and No.~DE-SC0011689; National Science Foundation,
Grant No.~0450696, and NSF-2013199; The Grainger Foundation; Marie
Curie-IRSES/EPLANET; European Particle Physics Latin American Network;
and UNESCO.
\end{sloppypar}

\end{document}